\documentclass[sigplan,10pt,nonacm]{acmart}
\pdfoutput=1
\renewcommand\footnotetextcopyrightpermission[1]{}
\def\BibTeX{{\rm B\kern-.05em{\sc i\kern-.025em b}\kern-.08emT\kern-.1667em\lower.7ex\hbox{E}\kern-.125emX}}

\usepackage{url}
\usepackage{xspace}
\usepackage[]{grffile}
\usepackage{color}
\usepackage{relsize}
\usepackage{algorithm}
\usepackage{algorithmicx}
\usepackage{multicol}
\usepackage[noend]{algpseudocode}
\usepackage{tikz}
\usepackage{amsmath}
\usepackage{amssymb}
\usepackage{datetime}
\usepackage{xcolor,listings}
\usepackage[T1]{fontenc}
\usepackage{csvsimple}
\usepackage{filecontents}
\makeatletter                   %1
\def\mdseries@tt{m}             %1
\makeatother                    %1
\usepackage[draft]{minted}
\usepackage{enumitem}
\usepackage[utf8]{inputenc}
\usepackage{subcaption}
\usepackage[htt]{hyphenat}

\usepackage{mdwlist}
\usepackage{enumitem}

\usepackage{makecell}

\newcolumntype{C}[1]{>{\centering\let\newline\\\arraybackslash\hspace{0pt}}m{#1}}
\definecolor{light-gray}{gray}{0.95}
\definecolor{darkred}{rgb}{0.8,0,0}

\hypersetup{
    colorlinks = true,
    linkcolor = purple,
    citecolor = {purple},
    urlcolor = {purple}
}

\newcommand{\system}{Weave\xspace}

\setcitestyle{square}

\date{}
\hypersetup{draft} 

\begin{document}

\title{Automating Cluster Management with Weave}
\author{Lalith Suresh, Joao Loff$^*$, Faria Kalim$^+$, Nina Narodytska, Leonid Ryzhyk, Sahan Gamage, Brian Oki, Zeeshan Lokhandwala, Mukesh Hira, Mooly Sagiv}
\affiliation{VMware, $^*$IST-Lisbon, $^+$UIUC}

\begin{abstract} % (fold)

Modern cluster management systems like Kubernetes and Openstack grapple with
hard combinatorial optimization problems: load balancing, placement,
scheduling, and configuration. Currently, developers tackle these problems by
designing custom application-specific algorithms---an approach that is proving
unsustainable, as ad-hoc solutions both perform poorly and introduce
overwhelming complexity to the system, making it challenging to add
important new features.

We propose a radically different architecture, where programmers drive cluster
management tasks declaratively, using SQL queries over cluster state
stored in a relational database. These queries capture in a natural way
both constraints on the cluster configuration as well as optimization
objectives. When a cluster reconfiguration is required at runtime, our
tool, called \system, synthesizes an encoding of these queries into an
optimization model, which it solves using an off-the-shelf solver. 

We demonstrate \system's efficacy by powering three production-grade systems
with it: a Kubernetes scheduler, a virtual machine management solution,
and a distributed transactional datastore. Using \system, we expressed
complex cluster management policies in under 20 lines of SQL, easily added
new features to these existing systems, and significantly improved
placement quality and convergence times.

\end{abstract}

\settopmatter{printfolios=true}
\maketitle

\section{Introduction}
\label{sec:introduction}

Modern cluster management systems like Kubernetes~\cite{kubernetes},
DRS~\cite{gulati2012vmware}, Openstack~\cite{openstack} and
OpenShift~\cite{openshift} are responsible for configuring a complex
distributed system and allocating resources efficiently. Whether juggling
containers, virtual machines, micro-services, virtual network appliances, or
serverless functions, these systems must enforce numerous cluster management
\emph{policies}. Some of these policies represent \emph{hard constraints},
which must hold in any valid system configuration, e.g., ``each container must
obtain its minimal requested amount of disk space'' or ``replicas of a service
must be placed on different server racks''. Others are \emph{soft
constraints}, e.g., ``scatter replicas across as many racks as possible''.
Given these constraints, a cluster manager must find a configuration
satisfying all hard constraints while minimizing violations of soft
constraints. 

Cluster manager developers usually tackle the problem by designing custom
application-specific algorithms---\emph{an approach that often leads to a
software engineering deadend} (\S\ref{sec:motivation}). As new types of
constraints are introduced, the developers get overwhelmed by having to solve
arbitrary combinations of increasingly complex constraints.  This is not
surprising given that, beyond the simplest cases, solving cluster management
constraints amounts to an \emph{NP-hard combinatorial optimization problem}
that cannot be efficiently solved via naive search.  In addition to the
algorithmic complexity, the lack of separation between the application state,
the constraints, and the constraint solving algorithm leads to unmaintainable
code. 

\begin{figure}[t] \centering
    \includegraphics[width=0.8\linewidth]{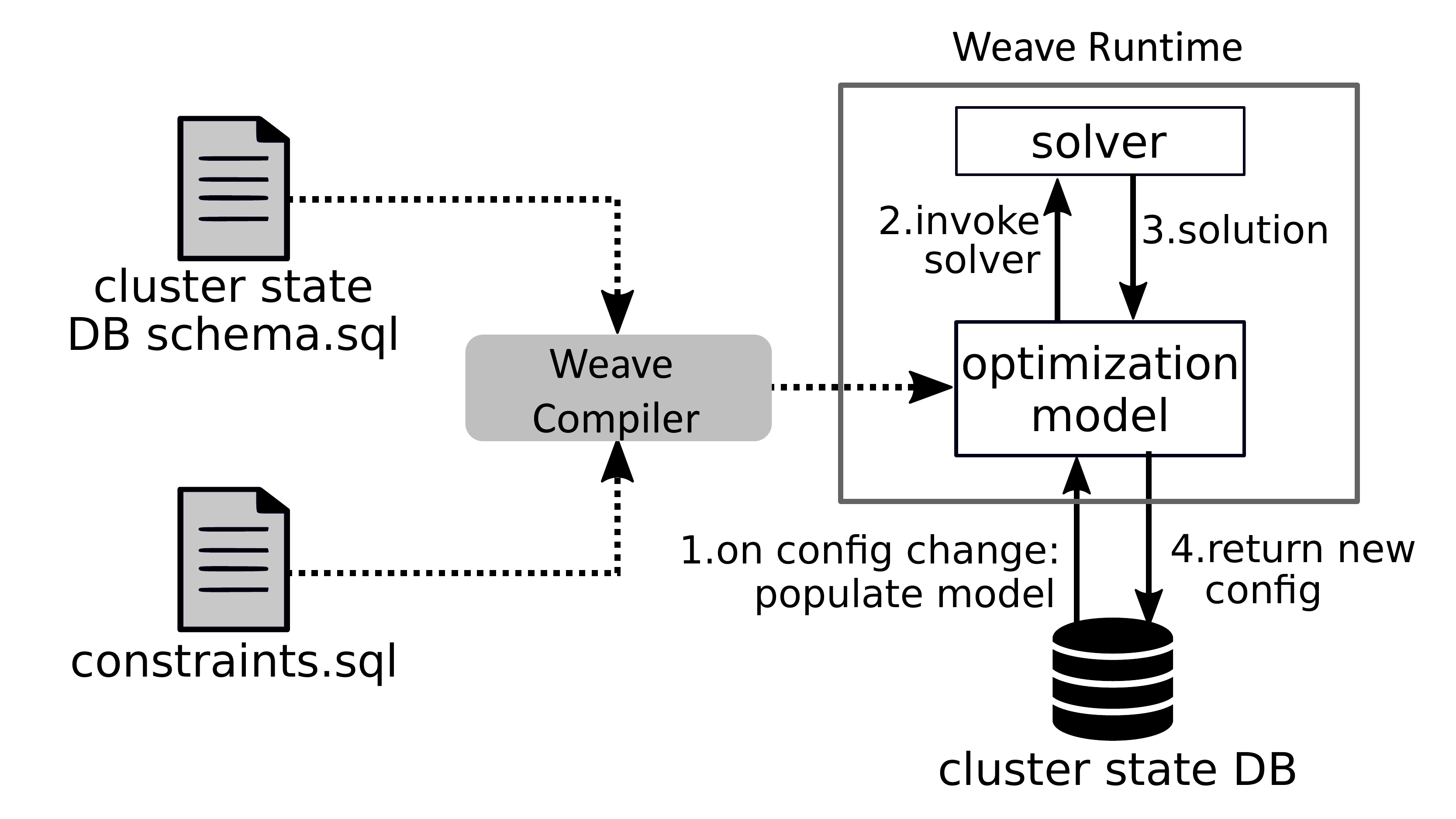}
    \caption{\system architecture. Dashed arrows show the model compilation flow.
    Solid arrows show runtime interactions between \system and the cluster state DB.} 
    \label{fig:weave_flow}
\end{figure}

To address these problems, we propose a radically different cluster manager
architecture and runtime framework, called \system (Figure~\ref{fig:weave_flow}).  With \system,
the developer maintains application state in a relational database, and
specifies constraints as database queries in SQL.  
Given this specification, \system compiles an \emph{optimization model} of the problem.
At runtime, when a system reconfiguration is required, \system populates the
model with the current state of the system extracted from the database,
solves it using an off-the-shelf solver, and generates a new configuration
that satisfies all model constraints.  By doing so, \system leverages decades of progress
on constraint solving technology, improves the separation of concerns in
the management plane, and frees the developer from the burden of maintaining an
ad-hoc solver.

%We propose a radically different management plane architecture and runtime
%framework, called \system, which disentangles the application state, policies and
%the solving algorithm (Figure~\ref{}).  With \system, the developer maintains
%application state in a relational database, and specifies policies as
%constraints on the state using the SQL language. At runtime, when a system
%re-configuration is required, \system combines these constraints with the current
%state of the system to formulate a \emph{combinatorial optimization} problem,
%which it solves using an off-the-shelf solver.  By doing so, \system leverages
%decades of progress on constraint solving technology, while improving the
%separation of concerns in the management plane and freeing the developer from
%the burden of maintaining an ad-hoc solver.

%\subsection*{Design principles}

%\label{sub:design_principles}

\system's design is based on three key principles:

\paragraph{1. A tool for engineers, not optimization experts}

The idea of solver-assisted system configuration has been explored by
researchers in the
past~\cite{Tumanov:2016:TGR:2901318.2901355,isard2009quincy,5934917,koster2010towards,narain2005network}.
However, effective use of solvers requires skills that most engineers do not
have, including formalizing system constraints as an optimization problem and
hand-crafting an efficient encoding of the problem using the language of the
solver.  As a result, solvers are rarely used in practice. 

In contrast, \system is designed to leverage existing languages, tools, and,
most importantly, existing expertise. With \system, developers use
off-the-shelf relational databases to manage cluster state as many systems
already do~\cite{openstack,vcenter}. In addition, they specify hard and soft
constraints on the cluster state using standard SQL.

\paragraph{2. Synthesize efficient optimization models}

Achieving good solver performance requires a carefully crafted problem
encoding, which in turn requires an in-depth understanding of solver
internals (\S\ref{sec:design}).  \system hides this complexity behind the high-level SQL syntax.
Internally, the \system compiler uses structural information extracted from
the SQL specification to synthesize an efficient encoding employing many
optimizations typically performed by expert users.

%\system instead \emph{generates} an efficient optimization model from the
%database schema and constraints written in SQL, that is then run by
%off-the-shelf solvers.

%As we will see in \S\ref{sec:design}, coming up with an efficient encoding for a
%solver requires an in-depth understanding of the solver's internals. \system hides
%this complexity from developers. \system instead \emph{generates} an efficient
%optimization model from the database schema and constraints written in SQL, that
%is then run by off-the-shelf solvers.

%At runtime, \system automates communication between the database and solver.
%When invoked, \system extracts the current cluster configuration from the
%database and converts it into an input to the solver.  It parses the solution
%computed by the solver and outputs the requisite modifications (deltas) to be
%made to the database state to satisfy all the policies. A controller, written
%by the developer, converts these deltas into lower-level control plane
%commands to the underlying system.

\paragraph{3. Exploit the incremental nature of cluster management}

Clever encodings go a long way in accelerating search for an optimal
configuration.  However, the fundamental NP-hardness of the problem kicks in
eventually, as the size of the system and the complexity of its constraints
grow.  This computational complexity barrier is an important concern in
data-center systems, which are expected to scale elastically to very large
workloads.

\system tames this complexity by leveraging the incremental nature of cluster management. In
production clusters, the workload and the physical topology change gradually
over time, with incremental changes, such as placing new tasks on the least
loaded hosts, often without touching existing workloads. From the optimization
perspective, restricting the scope of changes the system performs during
reconfiguration dramatically shrinks the search space explored by the
solver.

In \S\ref{sec:design_overview}, we show to use this capability to
build a Kubernetes scheduler that in the common case looks only at new
workloads that are awaiting placement. Only if it cannot find a feasible
solution, does it reason about all workloads in unison (and even
then the search is restricted to only perform a bounded number of changes to
the existing workload). In doing so, \system allows developers to
systematically deal with the computational complexity barrier.

\subsection*{Contributions}
\label{sub:contributions}

\hspace{0in} 
\indent    $\bullet$ We present \system, a system that helps the
    developers to automate complex cluster management tasks by using tooling
    familiar to most developers: the SQL language and relational databases.

    $\bullet$  We report in-depth about our experience building a Kubernetes
     Scheduler using \system: we not only implemented existing scheduling
     policies supported by Kubernetes, but also new policies in under 20 lines
     of SQL each. We also improved placement quality and convergence times
     over the baseline scheduler. We achieve all of these gains with a
     performance that is competitive with the highly-optimized baseline
     scheduler.

    $\bullet$ Lastly, we briefly describe our experience using \system to
    power two more systems: DRS where we significantly improved load balancing
    quality, and CorfuDB, where we implemented several cluster management
    features with a few lines of SQL.

% \end{itemize}

\section{Motivating Example} % (fold)
\label{sec:motivation}

\begin{figure}
\noindent
{\scriptsize
\begin{center}
        \begin{tabular}{@{}l|@{\,\,}p{2.8in}@{}}
         \textbf{Policy} & \textbf{Description} \\
         H1-4 & Avoid nodes with resource overload, unavailability or errors \\
         H5 & Resource capacity constraints: pods scheduled on a node must not exceed node's CPU, memory, and disk capacity \\
         H6 & Ensure network ports on host machine requested by pod are available \\
         H7 & Respect requests by a pod for specific nodes \\
         H8 & If pod is already assigned to a node, do not reassign \\
         H9 & Ensure pod is in the same zone as its requested volumes \\
         H10-11 & If a node has a `taint' label, ensure pods on that node are configured to tolerate those taints \\
         H12-13 & Node affinity/anti-affinity: pods are affine/anti-affine to nodes according to configured labels \\
         H14 & Inter-pod affinity/anti-affinity: pods are affine/anti-affine to each other according to configured labels \\
         H15 & Pods of the same service must be in the same failure-domain \\
         H16-20 & Volume constraints specific to GCE, AWS, Azure. \\
         S1 & Spread pods from the same group across nodes \\
         S2-5 & Load balance pods according to CPU/Memory load on nodes \\
         S6 & Prefer nodes that have matching labels \\
         S7 & Inter-pod affinity/anti-affinity by labels \\
         S8 & Prefer to not exceed node resource limits  \\
         S9 & Prefer nodes where container images are already available \\
        \end{tabular}
\end{center}
}
\vspace{-0.1in}
\caption{Policies from the baseline Kubernetes scheduler, showing both hard (H) constraints
and soft (S) constraints.}
\label{fig:policies}
\end{figure}

We study Kubernetes as a representative example of a modern cluster management
system and highlight the challenges it faces in solving the resource
management problem. These challenges are not unique by any means to
Kubernetes, and appear in many similar systems~\cite{openstack,borg,yarn}.

Specifically, we focus on the Kubernetes Scheduler, which is responsible for
assigning groups of containers, called \emph{pods}, to cluster nodes
(typically physical or virtual machines).  Each pod has a number of
user-supplied attributes that describe its resource demand (CPU, memory,
network, and storage) and placement preferences. The latter are typically
specified in terms of the pod's affinity or anti-affinity to other pods,
nodes, or groups of nodes. These attributes represent hard constraints that
must be satisfied for the pod to be placed on a node  (H1--H20 in
Table~\ref{fig:policies}). Kubernetes also supports soft versions of placement
constraints, with a violation cost assigned to each constraint (S1--S9 in
Table~\ref{fig:policies}).

Kubernetes solves these constraints using a greedy, best-effort, algorithm that
places one pending pod at a time.  To this end, it (i) selects all nodes that
satisfy the pod's hard constraints, (ii) ranks them based on soft constraint
violations, and (iii) schedules the pod on the top-ranked node.

%a growing list of constraints. It places one pod at a time according to a
%greedy, best-effort strategy: (i) the scheduler first filters out the set of
%valid nodes for the pod using a list of \emph{predicates} (hard constraints),
%(ii) it then ranks all the valid nodes using some priority functions (soft
%constraints), (iii) the pod is then placed on the highest ranked node.  The
%supported constraints enforce a range of scheduling policies, such as matching
%resource demands from pods to nodes, respecting affinity and anti-affinity
%requirements for pods, and load balancing.

This simple design suffers from three key problems.

%First, is the growing complexity of the code. And second, it is not guaranteed
%to find feasible placements, let alone find optimal placements.

\begin{figure}[t] \centering
    \includegraphics[width=0.9\linewidth, trim=30 50 0 0]{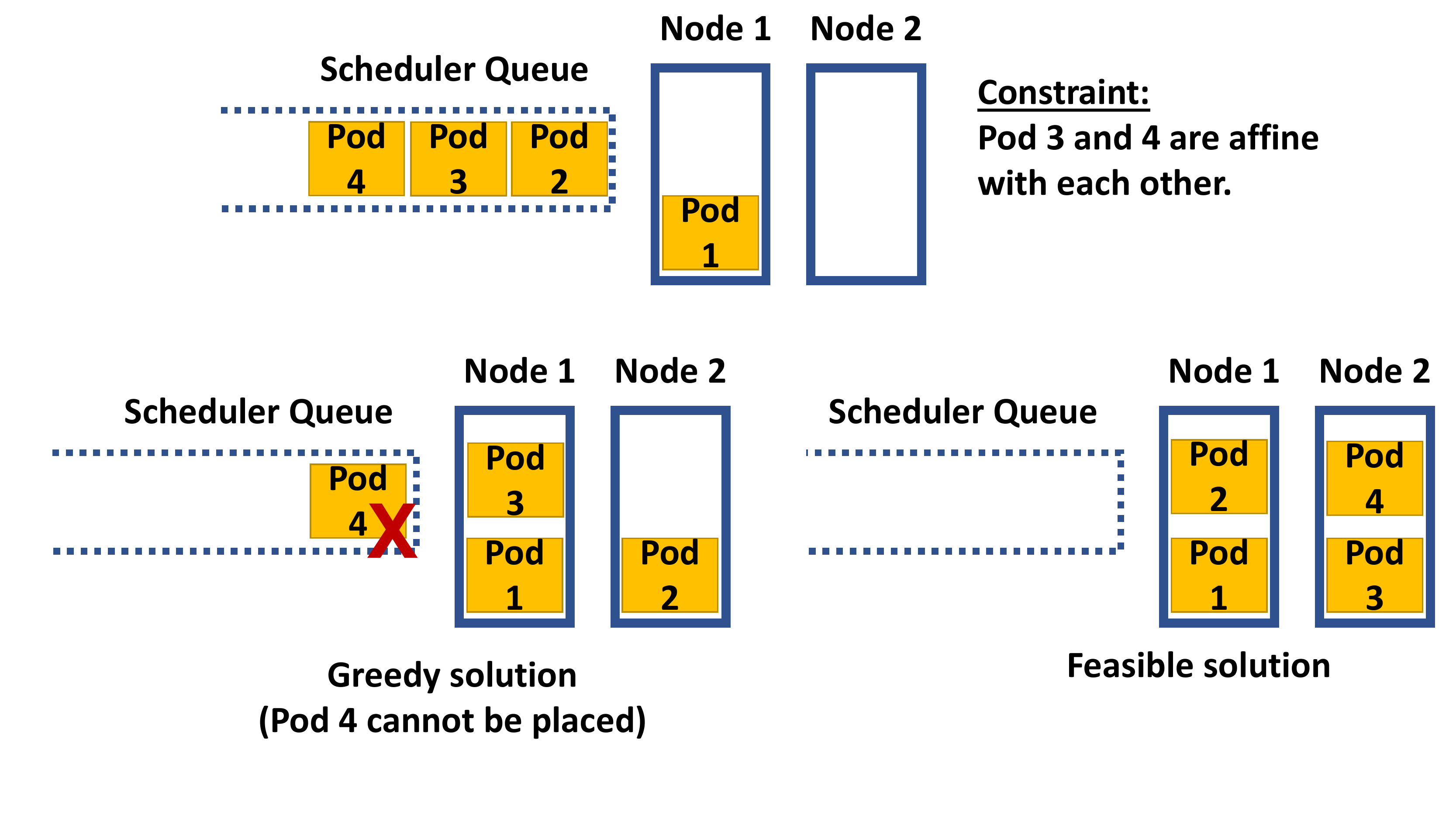}
    \vspace{-0.05in}
    \caption{An example with two nodes and three pods with different
    sizes. A greedy scheduler that places one pod at a time fails to place pod
    3, whereas a scheduler that places groups of pods at a time finds a
    feasible solution.}
    \label{fig:greedy_vs_batching}
\end{figure}

\vspace{-0.05in}
\paragraph{It is not guaranteed to find feasible, let alone optimal, placements}
Pod scheduling is a variant of the multidimensional bin packing
problem~\cite{albers2000average,bansal2006improved}, which is NP hard and
cannot, in the general case, be solved with greedy algorithms. This is
particularly challenging as the scheduling problem becomes \emph{tight}
because of more workload consolidation and users increasingly relying on
affinity constraints for performance and availability.
Figure~\ref{fig:greedy_vs_batching} shows an example with three pods
that the greedy algorithm fails to schedule: it places the first pod on the
least loaded node, leaving no room for the other two pods that must be placed
together on one node.

As a workaround, the scheduler attempts to preempt lower priority pods to make
room for incoming ones, which is both disruptive to running services and is in
itself not guaranteed to find feasible combinations of pods to preempt. In
Section~\ref{sub:k8s}, we quantitatively demonstrate that even when
applicable, pod preemption is slow to converge.

\begin{figure}[t] \centering
    \includegraphics[width=0.9\linewidth, trim=30 80 0 0]{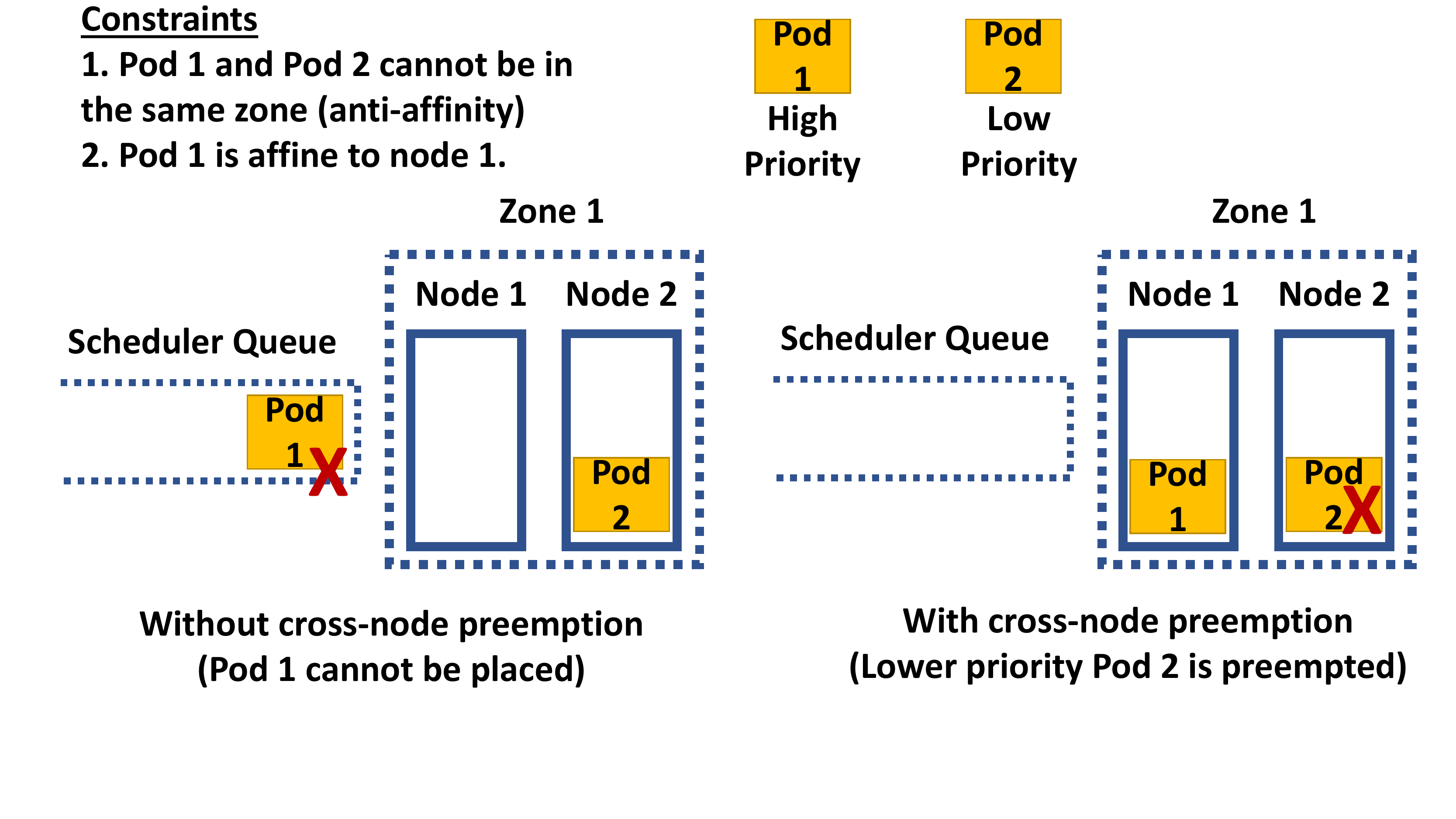}
        \vspace{-0.05in}
    \caption{An anti-affinity constraint prevents
    Pod 1 and Pod 2 from being in the same zone, pod 1 is affine to node 1,
    and pod 2 has a lower priority than pod 1. Placing pod 1 on node 1
    requires evicting pod 2 on node 2.}
    \label{fig:cross_node_preemption}
\end{figure}

\vspace{-0.1in}
\paragraph{Best-effort scheduling does not support global re-configuration}

Many scenarios require the scheduler to simultaneously reconfigure arbitrary
groups of pods and nodes.  For instance,
Figure~\ref{fig:cross_node_preemption} shows a scenario where a high priority
pod (pod 1) can only be placed on node 1, but to do so, the scheduler has to
preempt a lower priority pod on node 2. Computing this rearrangement requires
simultaneous reasoning about resource and affinity constraints spanning
multiple pods and nodes, which cannot be achieved within the current
framework.  Thus, although such global re-configuration is in high demand
among users, it is unsupported in
Kubernetes~\cite{crossnodepreemption,podAffinityHard}.

\vspace{-0.1in}
\paragraph{Best-effort scheduling leads to complex code}

%Greedy scheduling deals poorly with many useful classes of constraints,
%especially ones that are \emph{global} in nature (that is, the constraint
%applies over groups of pods and nodes). To efficiently evaluate such
%constraints, developers often introduce auxiliary data-structures, metadata and
%caching optimizations, which makes the code hard to maintain and evolve.

Greedy scheduling deals poorly with many useful classes of constraints,
especially ones that apply to groups of pods.  Consider Kubernetes'
\emph{service affinity} predicate. The predicate ensures that all pods from the
same service (e.g., a fleet of load balancers) are assigned to nodes that have
identical values for some label (e.g., an availability zone label). This
constraint holds over \emph{groups} of pods and nodes, but it is hard to
implement this in the current scheduler architecture that can only reason about
placing pods, \emph{one at a time}.  Therefore, the scheduler, when placing each
pod from a group, has to scan some data-structures to make sure its next
decision will be consistent with past decisions for pods from the same service.
The scheduler contains additional code to pre-compute these results to avoid the
inefficient scans for subsequent pods from the same group. Not surprisingly, the
code is commented with a large warning that indicates that the predicate is not
guaranteed to work without the precomputed metadata, and there is a discussion
to remove the feature because of the accumulating technical
debt~\cite{serviceAffinityHard}.

Similarly, there are discussions among developers about restricting
affinity/anti-affinity policies to make the code
efficient~\cite{podAffinityHard}.  Again, this challenge is aggravated by the
fact that a series of pre-computing optimizations used to speed-up such policies
are fragile in the face of evolving requirements (e.g., it is hard to
extend these policies to specify the number of pods per
node~\cite{podTopologyCount1,
podTopologyCount2,podTopologyCount3,podTopologyCount4,podTopologyCount5}).
The complexity accumulates making entire classes of policies
requested by users difficult to implement in the scheduler~\cite{crossnodepreemption,podAffinityHard}.

\section{\system by example} \label{sec:design_overview}

We show how \system automates cluster management by presenting a step-by-step
guide to building a Kubernetes scheduler with \system.  Our scheduler operates as
a drop-in replacement for the default scheduler (\S\ref{sec:motivation}),
supporting all of its capabilities and adding new ones.

The overall workflow of building a \system-powered scheduler consists of two
steps.  First, the developer stores the cluster state, including its current
and new workloads, in an SQL database. Second, the developer extends the
database schema with the specification of resource constraints, also written
in SQL.  The use of a cluster state database is inline with current practices,
as most existing cluster management systems rely on such a database as a
unified persistent representation of system state.  Thus in practice step one
is only needed if configuration is stored in a non-SQL database.

At runtime, our scheduler invokes \system when new pods are added to the system.
It detects cluster state database changes that require resource management
decisions to be made, solves the constraints and generates a new system
configuration as a database update (Figure~\ref{fig:weave_flow}).

\paragraph{Cluster state database} Kubernetes stores all state concerning nodes
and pods in an etcd~\cite{etcd} cluster.  The default scheduler maintains a cache of
relevant parts of this state locally using in-memory data structures.  To use
\system, we replace this cache with an in-memory embedded SQL database (Apache
Derby~\cite{derby}). As part of this process, the developer specifies an SQL schema
(tables and views) that represents the cluster's state. As of now, the schema
uses 18 tables and 9 views to describe the set of pods, nodes, volumes,
container images, pod labels, node labels and other cluster state.

The developer annotates some columns in the schema as \emph{decision
variables}, i.e., variables to be assigned automatically by \system. For
example, a placement decision of a pod on a node is represented by the table
in Figure~\ref{fig:pod_schema}. Here, the \texttt{@variable\_columns}
annotation indicates to \system that the \texttt{node\_name} column should be
treated as a set of decision variables. All other columns are \emph{input
variables}, whose values are supplied by the database.

\begin{figure}
\begin{minted}[fontsize=\footnotesize,
               frame=lines]{sql}
-- @variable_columns (node_name)
create table pending_pod
(
  pod_name varchar(100) not null primary key,
  status varchar(10) not null,
  namespace varchar(100) not null,
  node_name varchar(100),
  ... -- more columns 
);
\end{minted}
\caption{A table describing pods waiting to be scheduled. The \texttt{@variable\_columns} annotation indicates
that the \texttt{node\_name} column should be treated as a set of decision
variables. Other columns are input variables, whose values are supplied by the
database.}
\label{fig:pod_schema}
\end{figure}

\paragraph{Constraints} Next, the developer specifies constraints against the
cluster state as a collection of SQL views.  \system supports both hard and soft
constraints. Together, these constraints represent the cluster management \emph{policy}
that the developer aims to enforce.

\emph{Hard constraints} are specified as SQL views with the annotation
\texttt{@hard\_constraint}. For example, consider the constraint in
Figure~\ref{fig:hard_constraint_weave}, which states that \system is only allowed
to schedule pod $P$ on node $N$ if $N$ has not been marked unschedulable by the
operator, is not under resource pressure, and reports as being ready to accept
new pods.  This constraint is interpreted as follows: for all records returned
by joining the \texttt{pending\_pod} and \texttt{node} tables, the predicate
in the \texttt{where} clause must hold true.

\begin{figure}
\begin{minted}[fontsize=\footnotesize, 
               fontfamily=tt,
               frame=lines]{sql}
-- @hard_constraint
create view constraint_node_predicates as
select * from pending_pod
join node
  on pending_pod.node_name = node.name
where
  node.unschedulable = false and
  node.memory_pressure = false and
  node.disk_pressure = false and
  node.ready = true;
\end{minted}
\caption{A hard constraint to ensure pods
that are pending placement are never assigned to nodes that are marked
unschedulable by the operator, are under resource pressure, or do not
self-report as being ready to accept new pod requests.}
\label{fig:hard_constraint_weave}
\end{figure}

\begin{figure}
\begin{minted}[fontsize=\footnotesize, 
               frame=lines]{sql}
create view spare_capacity_per_node as
select (node.available_cpu_capacity 
        - sum(pending_pod.cpu_request)) as cpu_spare
from node
join pending_pod
  on pending_pod.node_name = node.name
group by node.name;

-- @soft_constraint
create view constraint_load_balance_cpu as
select min(cpu_spare) from spare_capacity_per_node;
\end{minted}
\caption{A soft constraint that maximizes the minimum spare CPU capacity
in the cluster for load balancing.}
\label{fig:load_balancing}
\end{figure}

\emph{Soft constraints} are also specified as SQL views. A soft constraint
view must contain a single record storing an integer value.  \system ensures
that the computed solution \emph{maximizes} the sum of all soft constraints.
All views with the annotation \texttt{@soft\_constraint} are treated as soft
constraints.

As an example, consider a load balancing policy to balance the CPU utilization
of nodes in a cluster (Figure~\ref{fig:load_balancing}). We first
write a convenience view (\texttt{spare\_capacity\_per\_node}) that computes the spare
CPU capacity after the pending pods have been placed.  We then describe a soft constraint view
(\texttt{constraint\_load\_balance\_cpu}) to compute the minimum spare capacity
in the cluster. By simply declaring such a view, \system computes solutions that
maximize the minimum CPU utilization of nodes in the cluster, thereby load
balancing pods across the cluster.

\emph{Challenging constraints:} Expressing more complicated policies than the
ones in Figures~\ref{fig:hard_constraint_weave} and~\ref{fig:load_balancing}
is just as straightforward using SQL.  Consider for example node affinity,
where a pod is affine towards nodes based on the values of the nodes' labels
(Figure~\ref{fig:node_affinity}).  Here, we specify that if a pod $p$ has node
affinity requirements, $p$ must be assigned to one of a set of nodes described
in another view \texttt{candidate\_nodes\_for\_pods} (which in turn enumerates
the set of candidate nodes for each pod according to node affinity rules).

All affinity and anti-affinity constraints we implemented have a similar
structure, with the latter using the \texttt{`not in'} operator instead of the
\texttt{in} operator. Views that involve only input columns such as
\texttt{candidate\_nodes\_for\_pods} are evaluated within the database,
whereas views involving variable columns are evaluated by \system within a
solver. This division of labor offloads a bulk of the complexity in
cross-referencing tables and computing joins to the database.

\begin{figure}
\begin{minted}[fontsize=\footnotesize,
               frame=lines]{sql}
-- @hard_constraint
create view constraint_node_affinity as
select *
from pending_pod
where pending_pod.has_requested_node_affinity = false or
      pending_pod.node_name in
         (select node_name
          from candidate_nodes_for_pods
          where pending_pod.pod_name = 
                candidate_nodes_for_pods.pod_name);
\end{minted}
\caption{A membership constraint to describe node affinity
(the pod must only be assigned to nodes it is affine to,
as computed in another view \texttt{candidate\_nodes\_for\_pods})}
\label{fig:node_affinity}
\end{figure}

\paragraph{At runtime,} whenever pods are added to the system, our Kubernetes
scheduler populates the \texttt{pending\_pod} table with new pods and invokes
\system to find an optimal placement for them.  By default, we schedule multiple
new pods at a time subject to a batching strategy.  As we will see in
\S\ref{sub:k8s}, this enables \system to find placements that respects pod
placement constraints and avoids redundant preemptions and scheduling failures
down the line. In the extreme case of using batches of size 1, our scheduler
degenerates to the behavior of the default scheduler.

% The scheduler also has an option to break up the pods into smaller batches,
% e.g., if scheduling them all at once takes too long, and schedule them
% separately by invoking \system once for each batch.  In the extreme case of using
% batches of size 1 our scheduler degenerates to the behavior of the default
% Kubernetes scheduler.  In practice, we never needed to split pods into batches
% in our experiments.

%and uses batching to reason about placing groups of pods at a time. It uses two
%optimization models to handle placement that differ mainly on the \emph{subset}
%of the cluster state that they operate upon. The first is an incremental model
%that only considers pods that have not yet been placed (that is, it operates on
%a view which is a tunable subset of the \texttt{pending\_pod} table). If that
%model fails to find a feasible placement, we revert to a model that looks at
%pods on a subset of nodes and preempts lower priority pods if required. Both
%models share most hard and soft constraints, with slight differences in some
%cases to accommodate the additional option of preemption in the second model.

When invoked by the scheduler, \system solves the constraints and outputs a copy
of the \texttt{pending\_pod} table with the \texttt{node\_name} column
assigned according to the computed optimal placement.  The scheduler then uses
this data to issue placement commands for each pod via the Kubernetes
scheduling API (the same API is used by the default scheduler).

\paragraph{Global reconfiguration} If no solution satisfying all hard
constraints exists, the scheduler attempts to find a more intrusive
reconfiguration that involves changing previous scheduling decisions.  Since
Kubernetes does not support pod migration, the only option available to the
scheduler is to evict one or more low-priority pods.  It therefore searches
for a solution that places a subset of new pods while possibly evicting some
existing lower-priority pods, minimizing the total number of evicted and
unscheduled pods.  This requires the developer to provide a slightly different
set of views and constraints that allow \system to change previous scheduling
decisions (see \S\ref{sec:design}).

%exists, \system generates an explanation of failure that identifies a minimal
%subset of constraints that cannot be satisfied simultaneously.  See
%Section~\ref{} for details.

%In both cases, \system outputs a table with variable columns that specifies pod
%to node assignments. Our scheduler uses the returned solution to issue the
%required control-plane commands in Kubernetes to place pods.

%\paragraph{Summary}
%Our example using Kubernetes illustrates several capabilities of \system. First,
%developers express complex cluster management policies using SQL in a modular
%fashion. These policies can be both hard constraints and soft constraints, and
%are natural to express using SQL. Second, \system synthesizes the required
%optimization model from the schema and constraints so that developers need not
%hand-craft efficient optimization models. Third, \system automates communication
%between the database and the solver at runtime. Finally, with \system,
%developers seamlessly control the search scope of solutions: a model can
%operate on an arbitrary subset of the cluster state to prevent model size and
%runtime blowups. In doing so, \system addresses both the adoption barrier and
%computational complexity barrier discussed in
%\S\ref{sec:introduction}.

\section{\system Design}
\label{sec:design}

\system enables programmers to specify cluster management policies using  a
high-level declarative language familiar to most programmers, and compute
policy-compliant configurations automatically and efficiently. Achieving this
however requires addressing some important challenges. 

First, given the amount of expertise required to hand-craft efficient
optimization models, it is non-trivial to bridge the gap between the SQL
representation of a problem and the formal modeling languages used by
optimization tools. We discuss how the \system compiler synthesizes efficient
optimization models in \S\ref{sub:synthesizing}.

Second, we describe how the \system runtime allows developers to easily tune
the search scope of a problem (\S\ref{ssub:runtime}).

Finally, given that programmers need systematic ways to test and debug the
policies that they write, we describe how we leverage a common solver
capability of finding \emph{unsatisfiable cores} to aid in debugging
(\S\ref{ssub:runtime}).

\subsection{\system compiler}
\label{sub:synthesizing}

The \system compiler generates an optimization model according to the database
schema and the constraints specified by the developer.

\paragraph{Syntax and expressiveness}

The compiler accepts SQL tables with integer, Boolean, and string columns.  It
supports a subset of the SQL query language, including its most commonly used
constructs: inner joins and anti-joins, aggregate queries, and correlated
sub-queries, as seen in Figures~\ref{fig:hard_constraint_weave} and
\ref{fig:load_balancing}.  It further supports arithmetic and logical
expressions built out of standard Boolean operators, linear arithmetic and
comparisons over integers, and equality checks over strings.  When using
backend solvers that support floating point types and non-linear arithmetic
(e.g., Gecode~\cite{gecode}), the compiler also allows these features in the
input specification.  Finally, the compiler supports a number of standard SQL
aggregate functions (\texttt{sum}, \texttt{min}, \texttt{max},
\texttt{count}), as well as additional aggregates that help in constraint
modeling.  These include, for example, the \texttt{all\_different} aggregate,
which enforces pair-wise inequalities among a set of variables
(\S\ref{ssub:generating_efficient_encodings}).  The compiler is
extensible, allowing user-defined aggregates to be added with a few lines of
code.

Using this syntax, we were able to compactly encode all hard and soft
constraints encountered in our case studies, including resource capacity,
affinity, anti-affinity, and load balancing constraints.  Formally, our
subset of SQL is sufficiently expressive to capture any constraints that are
expressible in the language of an ILP or constraint solver with no more than a
constant-factor blow-up in the size of the encoding.  In practice, SQL is much
more concise than the low-level languages supported by solvers, since a single
SQL view compiles into many low-level constraints.  For instance, the view in
Figure~\ref{fig:hard_constraint_weave} outputs one constraint per record in
the \texttt{pending\_pod} table.  In general, encodings synthesized by \system
are worst-case polynomial in the size of the cluster state database.

\paragraph{Intermediate representation} Internally, the compiler first converts
the program to an intermediate representation based on list-comprehension
syntax~\cite{Fegaras:2000:OOQ:377674.377676}, which makes it easy to apply
standard query optimization algorithms (like unnesting subqueries).  This
representation has been well-studied by the database
community~\cite{Grust96monoidcomprehensions,boag-xquery02,Fegaras:2000:OOQ:377674.377676,
Karpathiotakis:2016:FQO:2994509.2994516,karpathiotakis2015just,extendingXQuery}
and we therefore do not expand on it in this paper.

\paragraph{Backend}

The compiler backend generates an optimization model written using the
formal language supported by a specific solver, e.g., the language of linear
inequalities (for an ILP solver) or a constraint modeling language (for a
constraint solver).  The use of an IR facilitates support for multiple
backends, allowing system builders to easily try different types of solvers.
We have so far implemented a backend for the MiniZinc~\cite{minizinc}
constraint modeling language, which in turn supports a variety of solver
backends, and are implementing a backend for Google OR-tools~\cite{ortools}.

The model is parameterized by the content of the cluster state database,
encoded as model variables. At runtime, \system binds these variables to values
extracted from the database before dispatching the model to the solver.
Additional \emph{constrained variables} encode the optimization decision to be
computed by the solver to satisfy model constraints.

%With SQL being a high-level language compared to the native interfaces of
%different backends, Weave is also free to apply backend-specific optimizations
%during the translation phase. We discuss these details below.

\subsubsection*{Optimizing the optimization models} 
\label{ssub:generating_efficient_encodings}

\begin{figure}
\begin{subfigure}{.5\textwidth}
\begin{minted}[xleftmargin=20pt,
               xrightmargin=15pt,
               fontsize=\footnotesize, 
               highlightcolor=yellow!60,
               linenos,
               label=1) sum with filter,
               highlightlines={6, 7},
               frame=lines]{text}
% array of variables
array[1..1000] of var int: vars;
var int: num_values_above_ten = 
  sum([1 | i in 1..length(vars) where (vars[i] > 10)]);
\end{minted}
\label{fig:worst2}
\end{subfigure}%
\vspace{0.1in}
\begin{subfigure}{.5\textwidth}
\begin{minted}[xleftmargin=20pt,
               xrightmargin=15pt,
               fontsize=\footnotesize, 
               label=2) sum of predicate,
               highlightcolor=yellow!60,
               linenos,
               frame=lines]{text}
% array of variables
array[1..1000] of var int: vars;
var int: num_values_above_ten = 
  sum([(vars[i] > 10) | i in 1..length(vars)]);
\end{minted}
\end{subfigure}%
\vspace{0.1in}
\begin{subfigure}{.5\textwidth}
\centering
\begin{tabular}{rlrr}
  \hline
 & Encoding & Vars & Constraints \\
  \hline
1 & sum with filter & 6001 & 7001 \\
  2 & sum of predicate & 1000 &   0 \\
   \hline
\end{tabular}
\end{subfigure}%
\caption{Two equivalent MiniZinc encodings to compute the sum over a subset
of an array of variables. The specification only has one array of a 1000 variables
and no constraints. When compiled by the MiniZinc toolchain however, they result
in a dramatically different number of constraints and variables for the solver.}
\label{fig:minizinc}
\end{figure}

\begin{figure}
\centering
\begin{tabular}{rlrr}
  \hline
 & Encoding & Vars & Constraints \\
  \hline
1 & sum with filter & 40000 & 20100 \\
  2 & sum of predicate & 100 &   0 \\
   \hline
\end{tabular}
  \caption{Number of intermediate variables and constraints generated
  by the two equivalent approaches to encoding sums shown in Figure~\ref{fig:minizinc},
  but for the \texttt{demand\_per\_node} view computed in 
  Figure~\ref{fig:load_balancing} (for 100 pods and 50 nodes).}
\label{fig:minizinc_capacity_constraint}
\end{figure}

Solver performance is highly sensitive to problem encoding, with two
equivalent encodings often resulting in vastly different solver runtimes. The
\system compiler therefore uses backend-specific optimizations to generate
efficient models for the solver to evaluate. We discuss two of many
optimizations we perform to synthesize efficient models.

The MiniZinc toolchain we use for the following examples takes as input a
model in the MiniZinc language synthesized by the compiler.
At runtime, this model is populated with the content of the cluster state
database and translated it into a
lower-level representation called FlatZinc, which is finally dispatched to a
constraint solver. The number of variables and constraints in the FlatZinc program
directly affects the runtime and scalability of a given optimization model.

\paragraph{Re-writing to use fixed arity constraints} 
We discuss the problem of translating a seemingly straightforward SQL
operation---\emph{counting the number of elements in a column that are greater
than a constant}---into MiniZinc.  Consider the two MiniZinc statements shown in
Figure~\ref{fig:minizinc}. Both
these statements compute the number of elements in an array of 1000 variables
 whose value is larger than 10.

 In the first case, the solver cannot statically filter the list because the
 values of the variables (and therefore the arity of the list to sum) are not
 \emph{fixed yet}. This forces MiniZinc to internally make use of \emph{option
 types}, where a variable might be `absent'. The MiniZinc compiler
 translates option types into auxiliary variables and constraints in the
 generated FlatZinc model. For 1000 variables in the original array,
 this yields 7001 constraints and 6001 variables
 (Figure~\ref{fig:minizinc}).

 The second and less obvious model (Figure~\ref{fig:minizinc}) computes the \emph{sum of predicates},
 achieving the same result because the predicate
 evaluates to 0 or 1. However, it creates no additional variables
 or constraints, and the generated FlatZinc only contains the original 1000
 variables from the array. Note that this specific approach only works for
 summation; other aggregate functions need different optimizations.

Such optimizations have a drastic effect in complex models.
Figure~\ref{fig:minizinc_capacity_constraint} evaluates the impact of the same
optimization in compiling the \texttt{demand\_per\_node} view shown in
Figure~\ref{fig:load_balancing} for a topology with a hundred pods and fifty
nodes. Translating the SQL \texttt{where} clause using a filter in the
constraint program produces \emph{40K variables and 20.1K constraints} whereas
summing the predicates instead produces only 100 variables. Furthermore, the
runtime to execute the solver in each case reduces \emph{ten fold}, from 3.4
seconds to 0.32 seconds.

%Furthermore, the
%intermediate input file generated by Minizinc for the solver is 4.4MB in the
%first case (generated in 2.5 seconds) and only 4.7KB in the second case
%(generated in 250ms).

\paragraph{Re-writing to use global constraints}
\label{ssub:re_writing_expressions_to_use_global_constraints}

\begin{figure}
\begin{subfigure}{.5\textwidth}
\begin{minted}[xleftmargin=20pt,
               xrightmargin=15pt,
               fontsize=\footnotesize, 
               highlightcolor=yellow!60,
               linenos,
               label=Ensure all elements of vars are different,
               highlightlines={6, 7},
               frame=lines]{text}
constraint forall(i in 1..length(vars),
                  j in 1..length(vars) where j < i) 
                  (vars[i] != vars[j]);
\end{minted}
\label{fig:worst2}
\end{subfigure}%
\vspace{0.1in}
\begin{subfigure}{.5\textwidth}
\begin{minted}[xleftmargin=20pt,
               xrightmargin=15pt,
               fontsize=\footnotesize, 
               label=Global constraint,
               highlightcolor=yellow!60,
               linenos,
               frame=lines]{text}
constraint all_different(vars);
\end{minted}
\end{subfigure}%
\vspace{0.1in}
\begin{subfigure}{.5\textwidth}
    % \includegraphics[width=1\linewidth, trim=0 20 0 0]{figures-arxiv/variables_created_alldiff.pdf}
    % \begin{table}[ht]
      \centering
      \begin{tabular}{rlrr}
        \hline
       & Encoding & Vars & Constraints \\
        \hline
        1 & Without global constraint & 101 & 4951 \\
        2 & With global constraint & 101 &   2 \\
         \hline
      \end{tabular}
    % \end{table}
\end{subfigure}%
\caption{Two equivalent MiniZinc statements to enforce that all variables in an
 array take different values. The specification here has a 100 variables and
 a single constraint. With global constraints, the solver creates
 one additional variable and constraint each, whereas without using global constraints, the
 solver creates 4951 constraints.}
\label{fig:alldiff}
\end{figure}

Global constraints are constraints over \emph{groups} of variables for which
solvers implement specialized and efficient propagator algorithms that
dramatically reduce the search space of the problem. Consider for example the
common pattern of enforcing that a group of variables all take different
values. This pattern is especially useful in cluster management policies to
place replicas in different failure domains. 

One approach is to post this constraint as a list of pairwise
inequalities (Figure~\ref{fig:alldiff}, top). The other approach is to use the
\texttt{all\_different} global constraint (Figure~\ref{fig:alldiff}, bottom).
For an array of 100 variables, the former approach
instantiates 4950 integer constraints, whereas using a global constraint posts
only a single \texttt{all\_different} constraint, which in turn allows the
solver to use a specialized propagator.

\system identifies opportunities to transform common SQL patterns to use a suite
of global constraints. For example, we rewrite usages of the \texttt{IN}
operator in SQL and its arguments to instead use the \emph{membership} global constraint,
for which solvers typically implement fast propagators. \system also allows
developers to directly use global constraints, such as \texttt{all\_different} in SQL predicates.

\vspace{-0.08in}
\subsection{Interacting with the \system runtime}
\label{ssub:runtime}

\begin{figure}
\noindent
{\scriptsize
\begin{center}
        \begin{tabular}{@{}l|@{\,\,}p{4.6cm}@{}}
         \textbf{Operation} & \textbf{Description} \\ \hline
\texttt{model = \system.compile(schema)} & Invoke \system compiler to synthesize an optimization model from the SQL schema and constraints \\
\texttt{model.connect(db)} & Establish JDBC connection to the state DB \\
\texttt{model.solve(timeout)} & Solve constraints and return a set of tables \\  \hline
        \end{tabular}
\end{center}
}
\caption{\system interface}
\label{fig:interface}
\end{figure}

\paragraph{Interface} \system is packaged as a Java library that exposes the API
shown in Figure~\ref{fig:interface}.  The \texttt{compile()} method is
called once on startup.  It takes an SQL schema, describing cluster state
and constraints and invokes the \system compiler, which outputs a compiled optimization model
(\S\ref{sub:synthesizing}).  The \texttt{connect()} method is then called
to connect \system to an instance of the cluster state database.

The \texttt{solve()} method is called whenever the state database
changes, e.g., a pod is added to the system.
It populates the model with values extracted from the database (step~1 in Figure~\ref{fig:weave_flow}) and invokes
the solver to find an assignment to constrained variables (step~2).
\system converts the solution found by the solver (step~3) into a set of tables
returned to the caller (step~4).  The caller then reconfigures the
system according to this solution (this last step is not part of \system).

\system does not assume that the cluster configuration changes according to
the solution it computes.
The caller may choose not to apply the proposed reconfiguration,
the reconfiguration may fail completely or partially, cluster state
my change due to external events.  Therefore, \system does not memorize the
solution and instead treats the state database as the source of ground truth
about the system on every call to \texttt{model.solve()}.

\paragraph{Global reconfiguration}

If \system reports that problem constraints are unsatisfiable, the caller has three options: (i) bail out
and report the failure to the user, (ii) call \system again to solve a simplified version of
the problem, e.g., to schedule only a subset of pending pods, or (iii) attempt global
reconfiguration.  The last option adds more degrees of freedom to the problem, allowing \system
to search through reconfigurations that involve changing existing workload.

As discussed in \S\ref{sec:design_overview}, global reconfiguration requires a modified
SQL specification.  The necessary changes include annotating additional columns as variables,
as well as adding constraints to restrict values of these new variables.
Although the resulting specification shares the database schema and most constraints
with the original one, it is compiled as a separate model and initialized with its own
DB connection.  When the primary model fails to find a solution, the cluster manager
simply calls \texttt{solve()} on the second model to search for a global reconfiguration instead.

\paragraph{Testing and debugging models}

An important concern in using automatic cluster management systems like Kubernetes
is understanding why the cluster manager could not identify
a valid solution (e.g., a pod placement decision) in the face of a web of
constraints.  Did the cluster run out of resources?  Did the user choose an overly
strict set of constraints that cannot be satisfied simultaneously under current
CPU and memory pressure?  Or did the user mistakenly specify mutually contradicting
constraints, e.g., affinity and anti-affinity constraints over the same group of pods?
\system improves debuggability by taking advantage of a common
solver capability: that of identifying \emph{unsatisfiable cores}. An
unsatisfiable core is a minimal subset of model constraints and inputs that
suffices to make the overall problem unsatisfiable (for example, a core might
involve a single input variable that cannot simultaneously satisfy two
contradicting constraints in the model). 

We leverage this capability by providing a translation layer that extracts an
unsatisfiable core from the solver and identifies corresponding SQL
constraints and records in the tables that lead to a contradiction. We found
this invaluable when debugging complex scenarios involving affinity and
anti-affinity constraints. For example, a common pattern we experienced when
adding and testing new affinity policies was that the new policy
\emph{tightened the problem}, and therefore triggered a violation of some
other constraint in the system (such as resource capacity constraints). Rather
than suspect a bug in our specification of the new policy, the
unsatisfiable core rightfully points us to the contradiction between the
capacity constraint and the affinity requirement.

\section{Implementation}
\label{sec:implementation}

Our \system implementation is about 2K LoC in Java for the library and an
additional 1.5K LoC for tests. Our implementation uses the JOOQ
library~\cite{jooq} to conveniently interface with different SQL databases.
Our backend currently emits MiniZinc code. By using MiniZinc, we interface
with a variety of CP and MIP solvers; over the course of our study, we've used
Gecode~\cite{gecode}, Chuffed~\cite{chuffed}, Google OR-Tools~\cite{ortools},
and a commercial MIP solver as backends for \system. All our experiments
use the Google OR-Tools CP-SAT~\cite{ortools} solver, given that it is commercially
developed and performs well on MiniZinc benchmarks~\cite{minizincChallenge}.

Using MiniZinc introduces a trade-off between convenience and efficiency:
while MiniZinc is a high-level modeling language that is convenient to
generate, MiniZinc needs to combines the model \emph{and} the input data on
every invocation to produce FlatZinc code, which is what the solvers
interpret. This `flattening' step introduces additional latency (dominating as
much as 90\% of the solver invocation latency in our experiments). This
redundant flattening step is redundant and can be avoided by directly emitting
solver-specific code; we are currently working on a Google OR-Tools backend to
address this.

\vspace{-0.1in}
\section{Evaluation}
\label{sec:evaluation}

Our evaluation asks four key questions.

\begin{itemize}
    \item \textbf{Q1:} Is \system flexible enough to support diverse management use cases?
    \item \textbf{Q2:} Is it easy to write cluster management policies using \system?
    \item \textbf{Q3:} Is the synthesized solver code performant enough for real-world problem sizes?
    \item \textbf{Q4:} Can \system improve cluster management quality by using specialized solvers?
\end{itemize}

\noindent We answer all questions in the affirmative:

\textbf{Q1.} We present three case studies. We focus on the Kubernetes
    scheduler for depth (\S\ref{sub:k8s}). For breadth, we briefly describe
    case studies involving three other systems: DRS (\S\ref{sub:drs}) and
    CorfuDB (\S\ref{sub:corfudb}).

\textbf{Q2.}  We expressed a variety of cluster management policies across these studies.
    For Kubernetes, we implemented not only the policies supported by the
    baseline scheduler (\ref{ssub:expressiveness}), but also 3 additional
    policies that developers are
    currently struggling to implement efficiently
    (\S\ref{subsubsec:additional_policies}). All policies were
    implemented in less than 20 lines of SQL.

\textbf{Q3.}  Our Kubernetes scheduler is competitive with the highly optimized
    baseline scheduler with regards to throughput and scheduling latency (\S\ref{ssub:perf}). 

\textbf{Q4.} In the face of workloads with heterogeneous resource requirements,
    our Kubernetes scheduler has a higher success rate of placing all pods, 
    and is also twice as fast at converging for a workload
    that requires pod preemptions (\S\ref{ssub:perf}). For our virtual machine
    management use case, we significantly improve over the baseline load balancer
    (\S\ref{sub:drs}).

\subsection{Case study: Kubernetes} 
\label{sub:k8s}

We evaluate our implementation of a Kubernetes scheduler using \system. Like the
baseline scheduler, our scheduler runs as a pod within Kubernetes, and uses
the same REST APIs and hooks to consume and actuate upon the Kubernetes'
cluster state. Our scheduler comprises about 800 lines of Java code that acts
as a shim to \system, the large majority of which is the plumbing required to
store Kubernetes state in an in-memory SQL database. In addition, all the
tables, views and policies amount to \textasciitilde650 lines of SQL.

We evaluate both qualitative and quantitative gains in using \system to
implement the policies in the scheduler.

\subsubsection{How easy is to describe policies using SQL?} % (fold)
\label{ssub:expressiveness}

The Kubernetes scheduler has 20 configurable `predicates', which behave as
hard constraints, and 9 configurable `priorities', which are equivalent to
soft constraints (Table~\ref{fig:policies}). We implemented 15 of 20
predicates, where the predicates we skipped (H16--H20) were those that were
specific to GCE, AWS and Azure. We implemented all priorities as well. In
general, we found ourselves spending more time studying the existing
Kubernetes code and understanding the intent behind their policies than we did
writing the equivalent code in \system using SQL.

We were able to implement each of the constraints we considered using $<20$
lines of SQL. We count every SQL clause (\texttt{select}, \texttt{from},
\texttt{join}, \texttt{where}, \texttt{having} and \texttt{group by}) and
every predicate separated by and/or as a new line.  For example, the
\emph{service affinity} policy, discussed in \S\ref{sec:motivation}, which is
challenging to get right in Kubernetes, was implemented in only 6 lines of
SQL.  It merely involves joining one view (which itself is an additional 6
lines) and a table, and posting a constraint using a \texttt{having} clause.
As another example, a commonly used soft constraint in Kubernetes is the
\emph{least requested} policy, which load balances based on CPU and memory
utilization.  This policy was implemented in 4 lines of SQL -- we simply ask
\system to maximize the minimum sum of the CPU and memory slack per node.

The number of joins is a commonly reported complexity metric for SQL.  Except
for one constraint that involved two joins, all hard and soft constraints were
expressible using zero or one joins each.

In contrast, the existing Kubernetes scheduler consists of 10s of thousands of
lines of Go code that is deeply intertwined, and is challenging to maintain and
evolve (\S\ref{sec:motivation}).

\subsubsection{Scheduler performance characteristics} % (fold)
\label{ssub:perf}

We show that \system is competitive with the baseline scheduler
in terms of the time it takes to place pods in a cluster. We also present
lower-level measurements to show how \system behaves under the hood when making
placement decisions. 

\paragraph{Workload} 
The Kubernetes documentation describes a number of practical examples of
workloads involving a large and diverse collection constraints. These
workloads highlight capabilities of the current Kubernetes scheduler and thus
represent best-case scenarios for Kubernetes. We evaluate against such
workloads. We use a common pod affinity/anti-affinity pattern seen in
production workloads~\cite{podAffinityWorkload}, where nginx~\cite{nginx}
servers in a web application need to be co-located on the same machine as an
in-memory Redis cache~\cite{redis}.

In this scenario, each application requires: \emph{(i)} pods from the
in-memory cache to be anti-affine with one another, and are therefore placed
on different nodes, \emph{(ii)} the web-servers to be anti-affine with one
another as well, but to be affine with the cache pods. On a cluster of 50
nodes, we launch 30 such applications, for a total of 2400 pods. We measure
the time taken to complete all placements between the baseline scheduler and
\system. For \system, we measure under different batching granularities: we batch
pod requests for up to $b$ pods, or until 200ms have elapsed since the last
pod request was received. We test for $b=10$, $b=50$ and $b=100$.

\begin{figure}[t] \centering
    \includegraphics[width=1\linewidth]{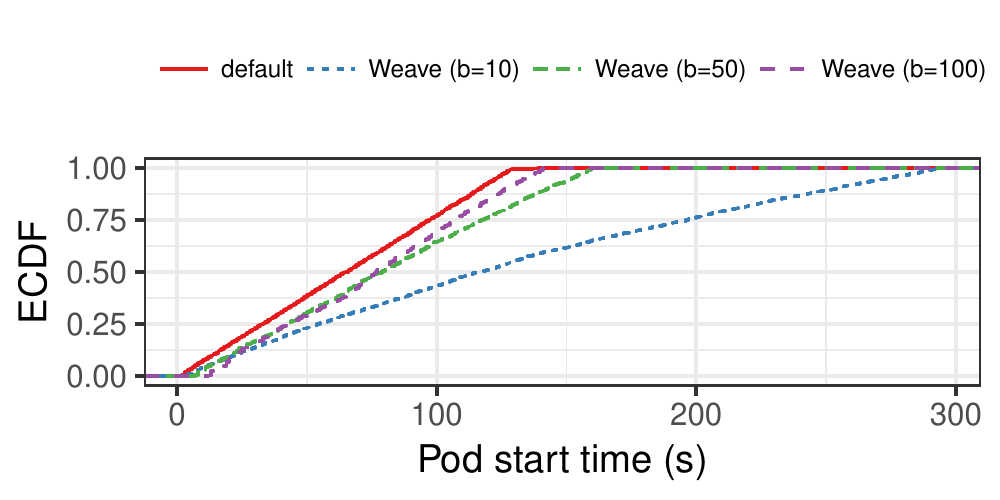}
    \caption{Pods placed over time for the affinity/anti-affinity workload. With
    enough batching, \system matches the baseline's rate of bringing up pods over time, as
    indicated by the identical slopes.} 
    \label{fig:pods_over_time_affinity_anti_affinity}
\end{figure}

\begin{figure}[t] \centering
    \includegraphics[width=1\linewidth]{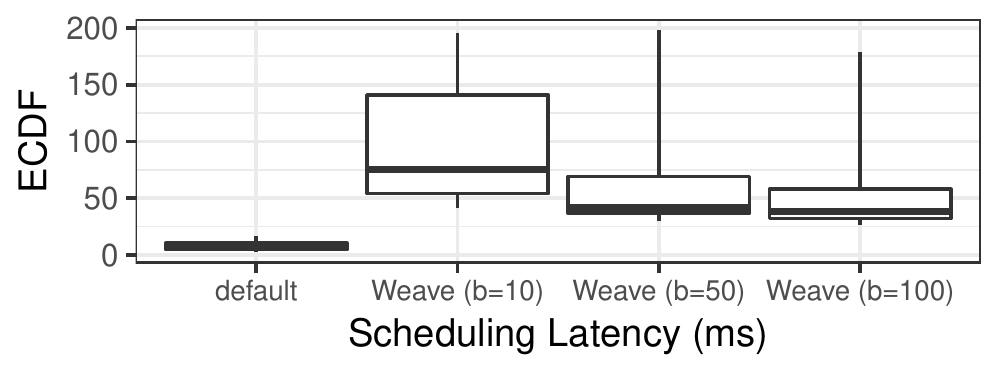}
    \caption{Scheduling latency, with boxplots and whiskers showing 5th, 25th, 50th, 75th and 95th
    percentile latencies for different configurations. At batch sizes of
    $b=100$ 50\% and 95\% of placement decisions are made in under 40ms and
    180ms respectively, which is small compared to the overall time taken to
    bring up a pod on a node (about 5 seconds).}
    \label{fig:task_placement_decision_time}
\end{figure}
\vspace{-0.1in}
\paragraph{Rate of pod placement} Figure~\ref{fig:pods_over_time_affinity_anti_affinity}
shows the fraction of pods (out of 2400) placed over time by i) the baseline
scheduler, and ii) \system with different batching granularities, $b$. With
enough batching ($b=50,100$), \system yields similar pod placement rates over
time as the baseline scheduler, as suggested by the identical slopes. Without
enough batching ($b=10$), \system does not amortize the cost of using a solver
well enough, causing pod placement latencies to increase (it takes 2.4x times
longer than the baseline). \system has a higher latency to place the first pod
because of the slow pod arrival rate and our use of batching (roughly 10 and
20 seconds for $b=50$ and $b=100$, consistent with the batching parameters);
this results in the gap seen between the curves for the default scheduler and
\system for $b=50$ and $b=100$.

\vspace{-0.05in}
\paragraph{Scheduling latency} Next, we measure the per-pod scheduling latency
incurred by \system. The latency for a pod is measured as the solving time per
batch divided by the batch size. This latency is measured over all the steps
required to invoke the solver when using \system: \emph{(i)} the time to prepare
the required input files for the solver, \emph{(ii)} converting the MiniZinc
program to FlatZinc, \emph{(iii)} running the solver, \emph{(iv)} parsing the
solution, and \emph{(v)} returning the computed solution back to the calling
program as a set of SQL records. Figure~\ref{fig:task_placement_decision_time}
shows our results. The baseline scheduler has a 95th percentile latency of
20ms. \system with enough batching makes 75\% and 95\% of placement decisions in
under 60ms and 175ms respectively, which is small compared to the overall time
taken to bring up a pod on a node (\textasciitilde5 seconds). Again, we see that
without enough batching ($b=10$), task placement latencies increase for \system
(75th percentile of 145ms).

The inefficient MiniZinc toolchain that we invoke out-of-process is the
primary cause of \system's additional latency, especially at the latency tail
when scheduling smaller batch sizes. In particular, the translation from
MiniZinc to FlatZinc incurs 25ms-1.5s of latency per batch in our experiments
depending on the problem size, accounting for as high as 90\% of the overall
latency of invoking a solver. This is however not fundamental to our approach
of using solvers; a \system backend that directly invokes the native interface
of a solver will not experience this overhead.

\begin{figure}[t] \centering
    \includegraphics[width=1\linewidth]{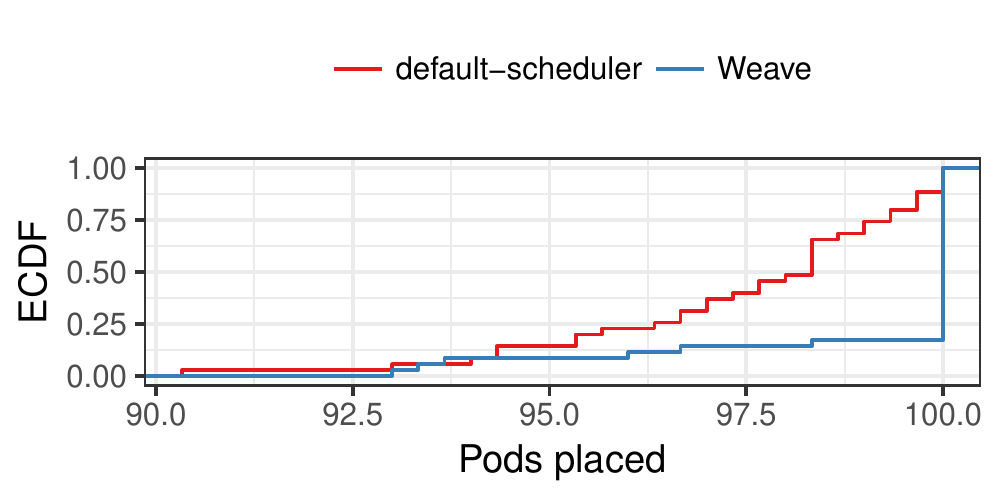}
    \caption{Placement quality: fraction of pods successfully placed by the default 
    and \system schedulers when pods have heterogeneous resource demands.} 
    \label{fig:placement_quality}
\end{figure}

\vspace{-0.05in}
\paragraph{Placement quality under heterogeneous workloads}
We now test how well the baseline and \system manage to find feasible placements
for all pods within a cluster (\S\ref{sec:motivation}). In our previous
experiments, the pods did not specify resource demands. In this experiment, we
test placement quality in the face of a realistic scenario where pods request
a mix of resource demands, which poses a harder scheduling problem. We repeat
the same workload as above for 30 applications each with 10 pods, but with pod
CPU and memory requirements following an exponential distribution. We generate
35 such workloads, which leads to a different arrival sequence of resource
demands per experiment. Figure~\ref{fig:placement_quality} shows the fraction
of pods that both \system and the baseline scheduler managed to place, across
all 35 runs. \system places 100\% of pods in 29 out of 35 experiments, and in
the worst case places at least 93\% of pods across all runs. In contrast, the
baseline scheduler packs all pods only in 3 out of 35 instances. This
highlights \system's effectiveness at placing \emph{groups} of pods. Instead,
the baseline myopically places one pod at a time, causing it to make decisions
that prevent future pods from being placed. Note, if the pods appear well
spaced apart in time, or \system uses smaller batching sizes, its performance
will approach that of the baseline.

\begin{figure}[t] \centering
    \includegraphics[width=1\linewidth]{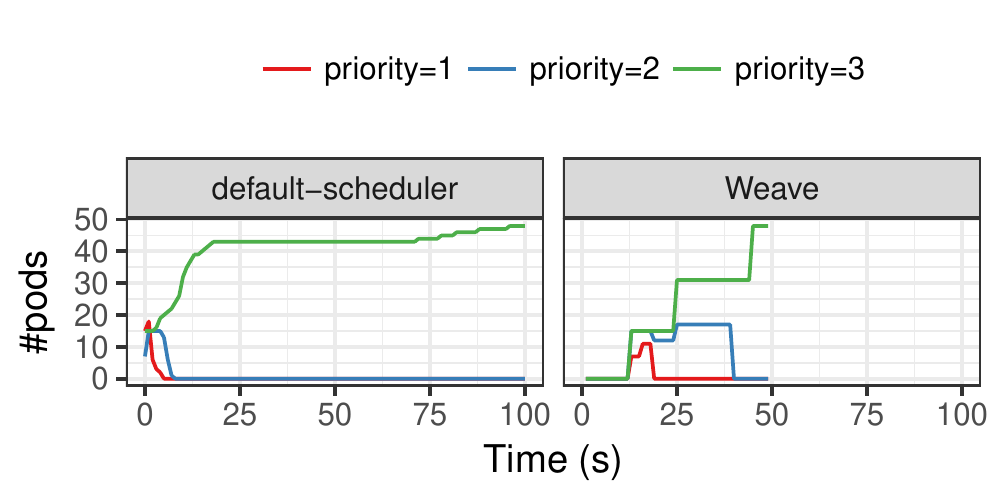}
    \caption{Preemption experiment: Timeseries showing number of pods from different priority levels
    over time.} 
    \label{fig:preemption}
\end{figure}
\vspace{-0.1in}
\paragraph{Placement convergence time for preemptions} We now test 
 \system's effectiveness in making global reconfiguration decisions. We replay a workload used
to test Kubernetes' preemption logic~\cite{podUnnecessaryPreemption}, that
creates 3 sets of pods with different priorities. The resource demands are set
to accommodate \emph{only} the highest priority pods on the cluster, and the
lower priority pods should either not be placed or be preempted.

Figure~\ref{fig:preemption} shows a timeseries depicting the number of pods
from each group (priority=3 are the highest priority pods). The baseline
scheduler invokes its preemption logic on a  pod-by-pod basis, and uses a set
of heuristics to determine when to retry pods it could not place (e.g
according to a backoff policy, and retrying when nodes report status updates).
In doing so, the baseline scheduler takes almost 100 seconds to place all high
priority pods. Instead, \system initially places low priority pods, and
systematically replaces them with higher priority pods in phases as new pod
requests arrive, by invoking its preemption model to look at the state of all
nodes (\S\ref{sec:design_overview}). In doing so, the scheduler converges to
placing all high priority pods in just under 50 seconds, twice as fast as the
baseline scheduler.

\begin{figure}[t] \centering
    \includegraphics[width=1\linewidth]{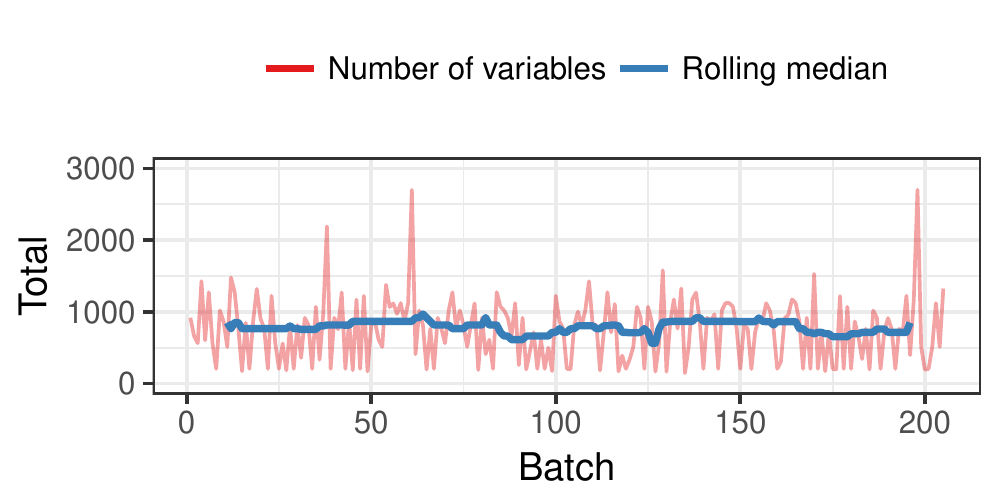}
    \caption{Number of variables per solver invocation over time, as the system grows
    from 0 to 2400 pods. The blue line shows a rolling median with a window of 20 points over the timeseries. The model size is constant w.r.t the system size.} 
    \label{fig:timeseries}
\end{figure}
\vspace{-0.05in}
\paragraph{Model size over time} To prevent model size blowups, \system allows
developers to bound the search scope of an optimization model and thereby
accommodate the incremental nature of cluster management. Our Kubernetes
scheduler, in the common case, only looks at pending pods that need placement,
and only reverts to a more global model when preemption is needed.
Figure~\ref{fig:timeseries} shows the size of the common case optimization
model in terms of the number of variables per solver invocation from one of
the runs that produced Figure~\ref{fig:pods_over_time_affinity_anti_affinity}.
While there is sufficient variability depending on the set of pods involved in
any scheduling batch, the model size trend (represented by a rolling median)
is near constant despite the growing number of pods in the system.

\vspace{-0.08in}
\paragraph{Optimizations under the hood} As discussed in \S\ref{sec:motivation}, by using
specialized solvers, we alleviate the need to carefully hand-craft caching and
pre-computing optimizations for efficiency, and instead, rely on sophisticated
algorithms within the solvers. For example, we noticed that even though the
FlatZinc input to the or-tools solver involved \emph{hundreds} of constraints
that capture anti-affinity requirements (constraints of the form $var_a \ne
var_b$), the solver identifies cliques of variables that are mutually not
equal, and rewrites them to instead use \texttt{all\_different} global
constraints which have highly efficient propagators (depending on the
composition of pod requests in the batch being processed, we noticed up to 16
\texttt{all\_different} constraints being created). Within a maximum of only
\emph{3ms per-batch}, the or-tools solver we used applies several such
pre-solving rules and re-writes the input model hundreds of times to simplify
the problem. In doing so, it exploits opportunities to simplify and speed up
solution search for an input problem that are otherwise challenging to
hand-craft and keep in sync as the system evolves.

\subsubsection{Implementing challenging policies}
\label{subsubsec:additional_policies}

We now add several new policies to Kubernetes. Each of these features
have been requested by users, are being discussed by developers, but are
challenging to implement efficiently.

\paragraph{Limiting the number of replica pods per node} Currently, 
affinity/anti-affinity constraints in Kubernetes control whether pods can or
cannot be co-located with one another. It does not allow users to limit the
\emph{number of pods} per service that can be co-located on a node. A typical
use case for such a policy is to limit the number of I/O intensive build tasks
from a CI server per machine~\cite{podTopologyCount4}.

Variants of this feature have been repeatedly requested by
users~\cite{podTopologyCount1,
podTopologyCount2,podTopologyCount3,podTopologyCount4,podTopologyCount5}, but
as the developers note, an efficient implementation for this conceptually
simple policy requires extensive and intrusive changes in the current code
base~\cite{podTopologyCount4}. However, implementing this capability using
\system required the same approach as posting the capacity constraints described
earlier: (i) a convenience view (8 lines of SQL) to compute for each replica
group that requires this constraint, the spare number of pods we can allocate
per node, and (ii) an additional 8 lines of SQL to post an aggregate
constraint per node and group. To test this policy, we mimicked the scenario
in Issue-40358~\cite{podTopologyCount4}, where we configured and deployed a
set of pods to have no more than 2 pods per node on our 50 node cluster.

\paragraph{Even pod spreading} In light of discussions around the previously
discussed policy, developers began work on a mechanism that enforces a hard
constraint to evenly spread pods across nodes or availability
zones~\cite{evenPodsSpreading1, evenPodsSpreading2}. However, this feature
cannot be added to Kubernetes in a modular fashion, and early patches involve
adding additional code paths within the affinity and anti-affinity logic.
Instead, this capability in \system merely involves 4 additional lines of SQL in
addition to the previously discussed policy (the limit on number of pods per
node is no longer a user-defined constant, but is $\left
\lceil{\frac{|pods|}{|valid\ nodes|}}\right \rceil$ or $\left
\lceil{\frac{|pods|}{|valid\ zones|}}\right \rceil$).

\paragraph{Cross-node preemption} Our preemption model is capable of global
reconfiguration as discussed in \S\ref{sec:design_overview}. By default, it
takes as input the set of all pods and nodes. It then reasons about packing as
many high-priority pods as possible by potentially preempting lower priority
pods, as shown in Figure~\ref{fig:preemption}. By simultaneously reasoning
about affinity and anti-affinity constraints across \emph{groups of pods and
nodes}, the scheduler is capable of identifying cross-node preemptions as
shown in Figure~\ref{fig:cross_node_preemption}.

\subsection{Case study: Virtual Machine Load Balancing} 
\label{sub:drs}

We now describe a tool we built for suggesting VM migrations in the context of
DRS~\cite{gulati2012vmware}. DRS has a highly optimized, greedy control
loop that handles online decisions. At slower timescales (e.g., every five
minutes), the system introspects the state of the entire cluster, and
identifies a series of VM migrations to make, with the objective of reducing
the overall standard deviation of node resource utilization (along multiple
resource dimensions).

\begin{figure}[t] \centering
    \includegraphics[width=1\linewidth]{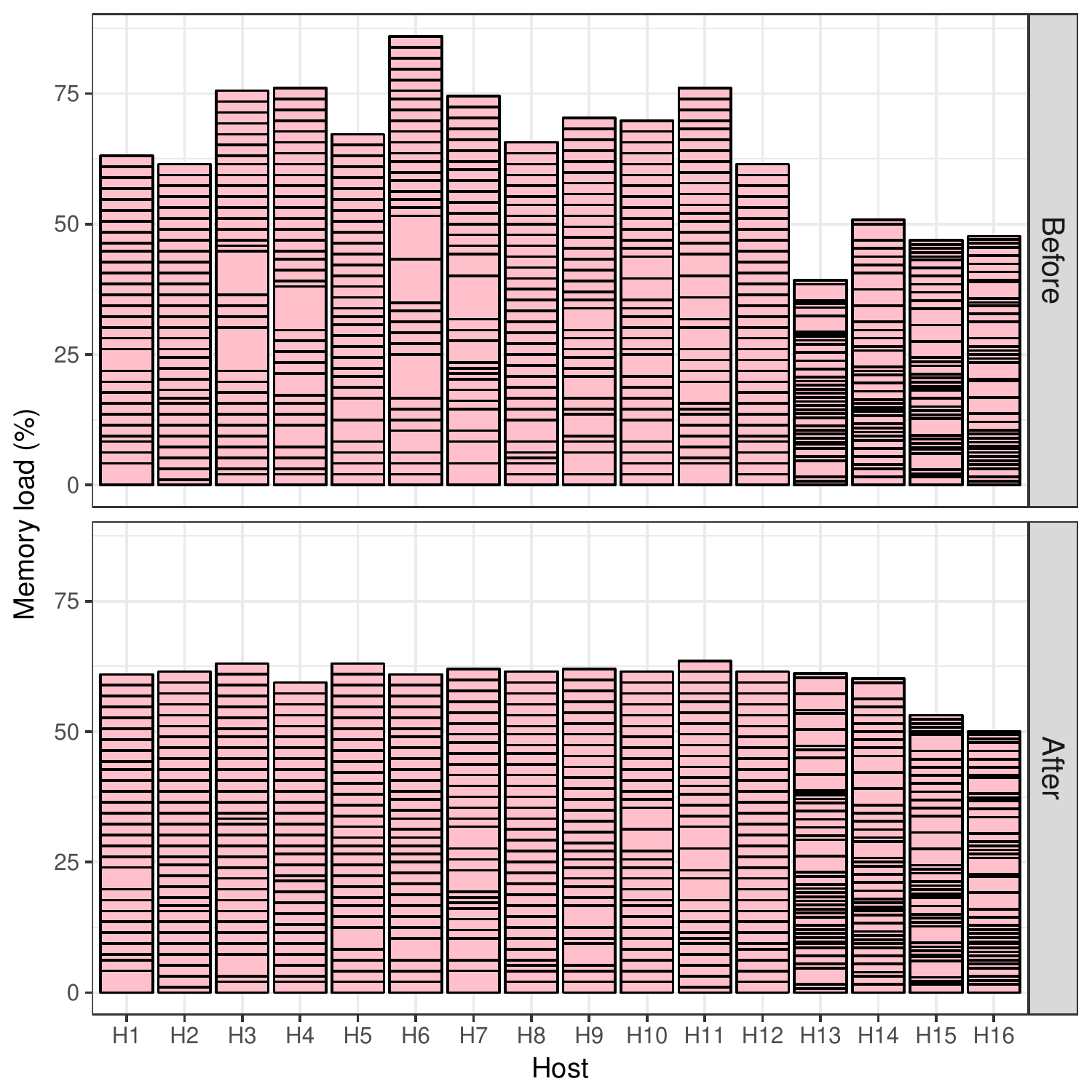}
    \caption{Load distribution before and after using \system. X-axis represents hosts,
    Y-axis represents the memory utilization per host. Hosts have
     different capacities, and boxes in each bar represent VM sizes,
    scaled to the host's capacity.}
    \label{fig:vm_load}
\end{figure}

We modeled the system in \system, and tested it using a trace from a bug report
submitted by a customer. This production cluster has 16 hosts with different
CPU and memory capacities, and 524 VMs with a range of CPU and memory sizes.
Figure~\ref{fig:vm_load} (before) shows the memory utilization of every host
as per the trace, where the boxes in each bar represent VM memory sizes,
scaled to the capacity of that host (there were no CPU resource reservations
by any of the VMs, so we do not show them). The baseline system could not
identify VM migrations to improve the load distribution of the cluster, which
led to the bug report.

With \system, we specified the necessary hard constraints (capacity and affinity
requirements) and a soft constraint that minimizes the load difference between
the most and least utilized node. We then made our tool identify ten VM
migrations, which resulted in the significantly more load balanced
distribution seen in Figure~\ref{fig:vm_load} (after). Previously, the most
loaded and least loaded nodes were at 85\% and 39\% utilization, whereas
afterwards, \system found an allocation where all nodes were between 51\% and
63\% utilization.

\subsection{Case study: Distributed Transactional Datastore} % (fold)
\label{sub:corfudb}

We implemented a management plane from scratch for a distributed data
platform~\cite{corfu} used in our production systems that supports
transactions. The system comprises a set of backend nodes that assume one or
more `roles' in the cluster. These roles include:
1) being an active transaction serialization server similar to the one used in
   Google Megastore~\cite{megastore} and Apache Omid~\cite{omid}, 2) being a backup
   serialization server in case the active fails, 3) data nodes that host
   several shards and replicas, which clients directly read/write from, and 4)
   a cluster of management servers that maintain the current cluster state
   using state machine replication. To integrate \system with this system, we modified the management
   servers to each maintain the state of the cluster in an embedded, in-memory
   SQL database.

We implemented multiple cluster management capabilities based on requirements
provided by engineers. We replicated some existing failure handling
policies, which decide when to remove a failed node from the cluster (as
opposed to marking it temporarily unavailable). We also added capabilities
currently unavailable in the system, like distributing roles across nodes,
and rack-aware placement of shards across data-nodes. All policies used
fewer than 10 LOC in SQL, as the cluster state was simple compared to
Kubernetes.

\vspace{-0.1in}
\section{Related work} % (fold)
\label{sec:related_work}
% \lalith{CITE ENTROPY PAPER}

% \lalith{Add B4, borg, autopilot}
% section related_work (end)

\paragraph{Use of solvers for resource management} A long body of work has
investigated the use of solvers for resource management including the use of
CP solvers~\cite{hermenier2009entropy} to pack and migrate VMs, flow network
solvers~\cite{isard2009quincy,199390} and MIPs for job
scheduling~\cite{Tumanov:2016:TGR:2901318.2901355}, and ILPs for traffic
engineering~\cite{danna2012practical}.
These systems expose the low-level language of the solver directly to
the user, requiring them to hand-craft their own optimization models.
In contrast, the \system hides
the complexity of using solvers behind the high-level SQL language and relational
databases, which is familiar to most developers.

Quincy~\cite{isard2009quincy} and Firmament~\cite{199390} use flow network
solvers for scheduling, which yield extremely quick solve times (sub-second,
even for topologies with thousands of nodes), but cannot model many
classes of constraints, including, e.g., inter-task constraints like
affinity/anti-affinity~\cite{GogIonelCorneliuFaec}. Furthermore,
they involve a high degree of modeling complexity: developers need to map
scheduling constraints and policies to primitive flow network constructs like
vertices, arcs, and flows, which makes it difficult to apply these tools to more general
cluster managers like Kubernetes. For example, the experimental Poseidon
Kubernetes scheduler~\cite{poseidon} is based on a flow network solver, but
the developers have found it hard to incorporate complex constraints for
pods~\cite{poseidonHard}.

As another example, Wrasse~\cite{Rai:2012:GRA:2391229.2391244} offers a DSL based
on the abstractions of balls and bins to specify resource allocation constraints and
a GPU-based parallel solver to quickly find solutions.  Similar to other systems in
this category, the low-level modeling language makes it hard to express complex
constraints; more importantly, it is restricted to resource capacity constraints,
but not, e.g., affinity, load balancing, and many other important cluster policies.

Some production systems use meta-heuristic search for resource management.
VMware DRS~\cite{gulati2012vmware} uses a greedy hill-climbing search, and
Service Fabric~\cite{serviceFabric} uses simulated annealing to perform
cluster resource management. To the best of our knowledge, neither
platform uses declarative programming techniques to describe and enforce
policies.

\vspace{-0.05in}
\paragraph{Network management} NetComplete~\cite{netcomplete},
SyNET~\cite{synet}, and Propane/AT~\cite{Beckett:2017:NCS:3062341.3062367}
synthesize configurations for specific network protocols like BGP and OSPF
from a high-level specification. Propane/AT and NetComplete use a custom DSL
to specify requirements and constraints, whereas SyNET uses Datalog. These
systems focus on configuration synthesis, which is orthogonal to \system, which
instead helps manage a dynamic cluster according to various constraints. That
said, the above-mentioned works involve path and routing constraints, given
their focus on protocols like BGP and OSPF. We have not explored such
constraints in our study.

ConfigAssure~\cite{configassure} and Alloy~\cite{narain2005network} use model
finding to identify configurations that satisfy a specification given by an
administrator (or detect errors in existing ones). Alloy uses a DSL for
specification, whereas ConfigAssure uses the Prolog language. \system
on the other hand works on top of standard SQL databases and is capable of
supporting optimization goals as well. The tested use cases for ConfigAssure
and Alloy are well within scope for \system.

Ravel~\cite{wang2016ravel} is an SDN controller that manages network state in
an SQL database, and uses database capabilities to both abstract and
orchestrate that state. \system also uses an SQL database to store cluster
state, but in addition uses a solver to \emph{search} for new configurations
based on constraints written in SQL.

% \paragraph{Relational databases for cluster state}
% Many production systems like
% OpenStack~\cite{openstack} and VMWare vCenter

% https://ratul.org/papers/pldi2017-propaneat.pdf
% https://www.usenix.org/system/files/conference/nsdi18/nsdi18-el-hassany.pdf

% smt solvers
% http://www.cs.princeton.edu/courses/archive/spring10/cos598D/FMINLecture5.pdf

% datalog-specific
% https://vanbever.eu/pdfs/vanbever_synet_cav_2017.pdf
% http://www.cs.princeton.edu/courses/archive/spring10/cos598D/FMINLecture5.pdf

\paragraph{DSLs for infrastructure automation}. Many tools use custom DSLs to
manage configurations. Hewson etal~\cite{hewson2012declarative} propose an
object-oriented DSL to specify a configuration for a data-center, which is
enforced by a constraint solver. PoDIM~\cite{delaet2007podim} does not use a
solver, but uses an SQL-like DSL to specify requirements for a configuration.
Configuration management tools like Puppet~\cite{puppet},
Ansible~\cite{ansible}, and Terraform~\cite{terraform} all use custom DSLs to
configure a fleet of servers. These systems target a different use case than
\system: they are not designed to operate within a dynamic distributed system
where the cluster state is constantly changing, and instead target
infrastructure deployment (which also runs at much slower timescales).

\section{Conclusion}
\label{sec:conclusions}

In this paper, we propose \system, a system with which developers can drive
cluster management tasks declaratively by specifying constraints using SQL, on
cluster state stored in a relational database. At runtime, during reconfiguration,
\system synthesizes these constraints into an optimization model that is
solved using an off-the-shelf solver.  We found it easy to apply \system
across a range of system management scenarios, which included building a
Kubernetes scheduler, to improving load balancing quality in a virtual machine
management solution, and to easily add features to a distributed transactional
data store.

\bibliographystyle{ACM-Reference-Format}
\bibliography{refs}

\end{document}